
%
%
\input harvmac.tex
\let\<=\langle
\let\>=\rangle
\let\e=\epsilon
\let\ep=\epsilon
\let\lam=\lambda
\let\tht=\theta
\let\t=\theta

\let\r=\rho
\let\l=\lambda
\let\p=\prime
\Title{\vbox{\baselineskip12pt\hbox{BUHEP-92-15}\hbox{YCTP-P13-1992}
   \hbox{hepth@xxx/9204094}}}
{\vbox{\centerline{N=2 Supersymmetry, Painleve III}
\vskip2pt\centerline{and Exact Scaling Functions in 2D Polymers}}}

\centerline{P. Fendley$^*$ and H. Saleur$^\dagger$}
\bigskip\centerline{$^*$Department of Physics}
\centerline{Boston University}\centerline{590 Commonwealth Avenue}
\centerline{Boston, MA 02215}
\bigskip\centerline{$^\dagger$Department of Physics}
\centerline{Yale University}
\centerline{New Haven, CT 06519}
\bigskip

\vskip .3in
We discuss in this paper various aspects of the off-critical $O(n)$ model in
two dimensions. We find the ground-state energy conjectured by Zamolodchikov
for the unitary minimal models, and extend the result to some non-unitary
minimal cases.  We apply our results to the discussion of scaling functions
for polymers on a cylinder.  We show, using the underlying $N$=2
supersymmetry, that the scaling function for one non-contractible polymer loop
around the cylinder is simply related to the solution of the Painleve III
differential equation.  We also find the ground-state energy for a single
polymer on the cylinder.  We check these results by numerically simulating the
polymer system.  We also analyze numerically the flow to the dense polymer
phase. We find there surprising results, with a $c_{\hbox{eff}}$ function that
is not monotonous and seems to have a roaming behavior, getting very close to
the values $81/70$ and $7/10$ between its UV and IR values of $1$.

\bigskip
\Date{April 1992}

\vfill\eject

Since the seminal paper on conformal field theory \ref\BPZ{A.A.Belavin,
A.M.Polyakov, A.B.Zamolodchikov, Nucl. Phys. B241 (1984) 333.}, there has been
steady progress in understanding two-dimensional critical systems (see
\ref\ISZ{C.Itzykson, H.Saleur, J.B.Zuber, {\it Conformal Invariance and
Applications to Statistical Mechanics}, (World Scientific, 1988)}). Properties
at the critical point are by now (almost) fully controlled, and new
developments have dealt with the vicinity of the critical point, where the
ultimate goal is to compute all scaling functions.  Most of the interesting
models turn out to be integrable. Factorizabilty then allows one to find the
$S$-matrix exactly \ref\Zamo{A.B.Zamolodchikov, Adv. Stud. in Pure Math. 19
(1989) 1}, and use it to calculate the exact ground-state energy
\ref\ZamoI{Al.B.Zamolodchikov, Nucl. Phys. B342 (1990) 695}. Although many
scaling functions are known in principle, their explicit computation is a
formidable task \ref\SmirnovI{F.A.Smirnov, {\it Form Factors in Completely
Integrable Models of Quantum Field Theory}, (World Scientific, 1992)}.  This is
in contrast with the beautiful and much older case of the Ising model, where
it was shown \ref\Mccoy{T.T.Wu, B.M.McCoy, C.A.Tracy, E.Barouch, Phys. Rev.
B13 (1976), 316} that spin correlation functions follow from solutions of the
Painleve III transcendent equation and its generalizations. Recently however,
similar equations have been met when studying ground-state properties of $N$=2
supersymmetric field theories
\ref\Cecotti{S.Cecotti, C.Vafa, Nucl. Phys. B367 (1991) 359}.
Using the underlying $N$=2 supersymmetry in polymers
\ref\SalI{H.Saleur, preprint YCTP-P38-91} we are able here to make contact
with these results and to exhibit a scaling function in the polymer problem
that is given by Painleve III. Remarkably, this property is derived directly
from continuum considerations, and awaits a lattice derivation.

We start by considering the $O(n)$ model. Although results are expected to be
universal we refer for concreteness to its realization on the hexagonal
lattice \ref\nienhuis{B.Nienhuis, J.Stat. Phys. 34 (1984), 731}.  Although the
model is defined originally for $n$ integer, the high-temperature expansion
allows analytic continuation to arbitrary $n$. The partition function reads
\eqn\Onpartition{Z=\sum K^{\# \hbox{monomers}}n^{\# \hbox{loops}},}
where the sum is taken over all configurations of self-avoiding and
mutually-avoiding loops that can be drawn on the lattice, and each elementary
edge that belongs to a loop is referred to as a monomer. Although this
geometrical definition involves global properties, locality can be restored if
one considers instead {\bf oriented loops} with a weight
$\exp(+i\alpha/6)$ for each left turn, and $\exp(-i\alpha/6)$ for each
right turn. When $n=2\cos\alpha$, the relation $|\#\hbox{left turns}-\#
\hbox{right turns}|=6$ for a closed loop on a planar hexagonal lattice gives
the correct weight for each loop. (We discuss later what happens on a
cylinder.)

 The model has interesting phase transitions for $n$ between -2 and 2. The
standard (also called dilute) critical point occurs for $K_c =
[2+(2-n)^\half]^{-\half}$. Parametrizing
\eqn\Nmin{n=2\cos{\pi\over t},}
with $2<t<\infty$, the continuum theory at $K=K_c$ has central charge
\ref\Dots{V.Dotsenko, V.A.Fateev, Nucl. Phys. B240 (1984) 312}
\eqn\cdilute{c = 1 - 6/t(t+1).}
The phase $K<K_c$ is disordered. The phase $K>K_c$ is massless and is referred
to as dense phase, because the polymers cover a non-vanishing fraction of
space even in the continuum limit. The continuum limit does not depend on
the value of $K$, and is a conformal field theory with central charge
\eqn\cdense{c= 1 - 6/(t-1)t.}

In the dilute critical theory the thermal operator (i.e.\ the one coupled to
$K$) is described as $\Phi_{13}$ with conformal weight $h={t-1 \over t+1}$. As
discussed in \ref\ZamoII{A.B. Zamolodchikov, Mod. Phys. Lett. A6 (1991) 1807}
this perturbation is integrable.  We discuss mainly the case $K>K_c$ in the
following. The corresponding S-matrix has been studied in \ZamoII; the
integrability requires that the $N$-body S-matrix elements factorize into a
product of two-body ones and thus obey the Yang-Baxter equation.  One starts
by considering the $O(n)$ model for $n$ integer and assumes that there is a
particle for each color, whose trajectories may be thought of intuitively as
the loops in \Onpartition. The $O(n)$ symmetry allows the S-matrix to be
written in terms of a set of invariant tensors, where these tensors obey an
algebra depending on $n$. Using this algebra, the Yang-Baxter equation reduces
to a single functional equation with scalar quantities, where $n$ appears as a
parameter.  Analytic continuation becomes possible, although for $n$
non-integer it does not define a bonafide S-matrix. However, quantities we
calculate using this S-matrix should be analytic in $n$. Therefore, the
form-factors calculated in \ZamoII\ and the ground-state energy levels derived
here are valid for any $n$.

As was also observed in \ref\Smir{F.A.Smirnov, Phys.Lett. B275 (1992) 109} in
a slightly different language, the main property of the S-matrix of \ZamoII\
is that the invariant tensors obey the Temperley-Lieb algebra
\ref\Ppm{P.Martin, {\it Potts Models and Related Problems in Statistical
Mechanics}, (World Scientific, 1991)} (with parameter $n$). As in lattice
models, many interesting properties do not depend on the particular
representation of this algebra. Thus we can switch from the representation
used in \ZamoII, defined only for $n$ integer, to the 6-vertex model
representation \ref\Bax{R.J. Baxter, {\it Exactly Solved Models in Statistical
Mechanics}, (Academic Press, N.Y., 1982)}, defined for any $n$. In the present
context, this amounts to considering the sine-Gordon S-matrix
\ref\ZamoIII{A.B.Zamolodchikov, Al.B.Zamolodchikov, Ann. of Physics 120 (1979)
253}
\eqn\Sinegordon{S={Z(\tht)\over\sinh\left({\tht-i\pi\over t}\right)}
\left[\sinh\left({\tht-i\pi\over t}\right)
-\sinh\left({\tht\over t}\right)e\right],}
where
\eqn\ematrix{e=q^{-1}E_{11}+qE_{22}-E_{12}-E_{21},}
$q=e^{i\pi/t}$, $(E_{ij})_{kl}=\delta_{ik}\delta_{jl}$.  In this new picture
the massive particles are the sine-Gordon soliton-antisoliton pair whose
trajectories can be thought of as the oriented loops in the local formulation
of the $O(n)$ model. The S-matrix \Sinegordon\ differs from the usual one by a
gauge transformation that ensures proper algebraic properties\nref\Leclair{
A.LeClair, Phys. Lett. B203 (1989) 103}\nref\Smirnov{ F.A.Smirnov, Nucl. Phys.
B337 (1990) 156} \refs{\Leclair,\Smirnov}, but our calculation is invariant
under this transformation.  There are no bound states in the spectrum for the
sine-Gordon couplings we will study.

We shall derive in the following the ground-state energies $E(R)$ of the
$O(n)$ model on a cylinder of radius $R$ and large length $L$. We will
calculate these energies for a variety of boundary conditions around the
cylinder. It is well known \ref\PasSal{V.Pasquier, H.Saleur, Nucl. Phys.  B330
(1990) 523} that when one switches from the $O(n)$ S-matrix to the sine-Gordon
S-matrix, boundary conditions must be treated very carefully. This is similar
to the situation in the lattice model when one switches from the geometrical
definition to the local one involving oriented loops. On a cylinder, there are
non-contractible loops in the $O(n)$ model, and for these the number of left
and right turns are equal \ref\Dfsz{P.Di Francesco, H.Saleur, J.B.Zuber, J.
Stat. Phys. 34 (1984) 731}. In the local formulation they therefore get a
weight 2 instead of the desired value $n$.  This boundary condition does not
exactly reproduce the $O(n)$ model; it amounts to introducing at infinity an
operator of conformal dimension $h=-1/4t(t+1)$
\ref\Isz{C.Itzykson, H.Saleur, J.B.Zuber, Europhys.  Lett. 2 (1986) 91}.

A useful check on the results is to to take the ultraviolet (mass
$\rightarrow 0$) limit of $E(R)$, because a general result of conformal field
theory \nref\rBCN{H.Blote, J.Cardy, M.P.Nightingale, Phys. Rev. Lett.
56 (1986) 742}\nref\rAff{I. Affleck, Phys. Rev. Lett. 56
(1986) 746} \refs{\rBCN,\rAff} gives
\eqn\eBCN{E_{UV}=-{\pi\over 6 R}(c-12h-12\bar h),}
where $h$ and $\bar h$ are the conformal dimensions of the operator creating
this state. The ground state is that created by the lowest-dimension operator;
only in unitary theories does this have $h=\bar h=0$. In non-unitary theories
one often defines an effective central charge $c_{\hbox{eff}}$ as
 $-6RE(R)/\pi$.  Thus in the case where non-contractible loops have weight $2$,
$c_{\hbox{eff}}=1$ in the UV. We will show that, as expected, the behavior of
$E(R)$ under thermal perturbation is given by that for the sine-Gordon model
with unmodified boundary conditions, which indeed has $c_{UV}=1$.

The technique we will use to calculate the ground-state energies from the
S-matrix is known as the thermodynamic Bethe ansatz (TBA) \ZamoI. The TBA
gives the free energy $f=-T \ln Z$ of a particle gas on a line of large length
$L$ at a temperature $T$. This is equivalent to considering a spacetime of a
cylinder of radius $R=1/T$ and length $L$. If we change our point-of-view and
think of the spatial direction as (Euclidean) time, only the ground state
contributes to $Z$, since only the lowest-energy state can propagate over the
large distance $L$. Thus $Z=\exp(-LE(R))$, where $E(R)$ is the ground-state
(Casimir) energy with space a circle of radius $R$. The equivalence between
the two approaches (often called modular invariance) means that $E(R)=-\ln
Z/L=R f/L$.  Introducing chemical potentials allows one to change the
boundary conditions in the $R$ direction \ref\fend{P.  Fendley,
``Excited-State Thermodynamics,'' to appear in Nucl. Phys. B.}.  The TBA
equations for arbitrary chemical potential \ref\KM{T. Klassen, E. Melzer,
Nucl. Phys B350 (1990) 635} give the ground-state energy
\eqn\fe{E_{\{\l\}}(R)=-\sum _a {m_a\over 2\pi }\int d\t
\cosh \t \ln (1+\l_a e^{-\e _a(\t)}),}
where the $\e_a$ obey the integral equations
\eqn\TBA{\e _a(\t )=m_aR\cosh (\t )-\sum _b \int {d\t ^{\p}\over 2\pi}\phi
_{ab}(\t -\t ^{\p})\ln(1+\l_b e^{-\e _b(\t ^{\p})}).}
The index $a$ labels the particle species, $\l_a$ are fugacities, $m_a$ are
the particle masses and the $\phi_{ab}$ depend on the S-matrix. If the
S-matrix is not diagonal, some of the particles in \TBA\ are zero-mass
``pseudo-particles'', which result from the diagonalization of the S-matrix.

\nref\Conj{Al.B.Zamolodchikov, Nucl. Phys. B358 (1991) 497}
\nref\FI{P. Fendley, K. Intriligator, Nucl. Phys. B372 (1992) 533}
To do the TBA calculation, we need to diagonalize the sine-Gordon S-matrix,
meaning that we find the eigenvalues $\Lambda(\tht_i| \tht_1,
\tht_2,\dots,\tht_N)$ for bringing the $i$th particle of momentum
$p_i=m\sinh\tht_i$ through all the others \refs{\Conj,\FI}. These eigenvalues
are defined precisely in the Appendix. Diagonalizing the sine-Gordon transfer
matrix is formally equivalent to diagonalizing the XXZ model transfer matrix,
so doing our thermodynamics is very similar to doing the thermodynamics of the
XXZ model, which was considered in \ref\TS{M.Takahashi, M.Suzuki, Prog. Theor.
Phys. 48 (1972) 2187}.  Putting periodic boundary conditions in the $L$
direction quantizes the $i$-th momentum via
\eqn\quant{e^{ip_iL}\Lambda(\tht_i| \tht_1, \tht_2,\dots,\tht_N)=1.}
This enables us to define the density of states $P_0(\tht)$ for the particles,
whether solitons or antisolitons (since the scattering is elastic, the total
number of particles is conserved). Taking the logarithm of \quant\ gives
$$2n\pi=m\sinh\tht_i+Im\ln \Lambda(\tht_i|\tht_1,\dots,\tht_N)$$ where $n$ is
an integer. Then
\eqn\dos{2\pi P_0(\tht_i)=2\pi {dn\over d\tht_i}=
mL\cosh\tht_i+{d\over d\tht_i}
Im\ln \Lambda(\tht_i|\tht_1,\dots,\tht_N).}
To find the partition function, one takes the thermodynamic trace over all the
eigenstates, which is equivalent to summing over all soliton and antisoliton
configurations. The difficulty of this study depends on arithmetic properties
of the parameter $t$.

We shall first consider the case $t$ integer. In this case, the eigenvalues
for the XXZ transfer matrix can be written as functionals of densities of
pseudoparticles called $a$-strings with $a=1,2,...,t-1$ and an
antipseudoparticle $(1^-)$.  Call $P_a^+$ the associated densities, $P_a^-$
the corresponding densities of holes, and $P_a=P_a^+ + P_a^-$. Call also
$P_0^+$ the density of real particles---the states which are actually occupied
by solitons or antisolitons; the density of corresponding holes $P_0^-=P_0
-P_0^+$.  As derived in \FI\ in the case $t=2$ and in the general case in the
Appendix, the densities obey the equations
\eqn\lastconstr{2\pi P_0(\t)=mL\cosh\t+\phi_{12}*\rho_1,}
and
\eqn\diagr{2\pi P_a=\sum_b \phi_{ab} * \rho_b,}
where $\phi _{ab}(\t )=l_{ab}(\cosh (\t ))^{-1}$,
$\r_0=P_0^+,\r_{1^-}=P_{1^-}^+$, and $\r_a=P_a^-$ otherwise; $*$ denotes
convolution.  The sum is over all particles, real and pseudo; $l_{ab}$ is the
incidence matrix for the diagram
\bigskip
\noindent
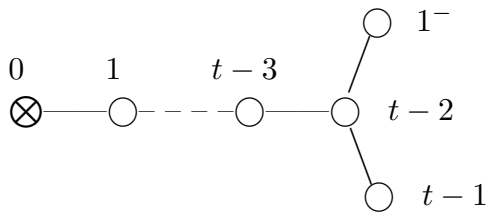
\centerline{\hbox{\rlap{\raise28pt\hbox{$\hskip4.5cm\bigcirc\hskip.25cm 1^-$}}
\rlap{\lower27pt\hbox{$\hskip4.4cm\bigcirc\hskip.3cm t-1$}}
\rlap{\raise15pt\hbox{$\hskip4.1cm\Big/$}}
\rlap{\lower14pt\hbox{$\hskip4.0cm\Big\backslash$}}
\rlap{\raise13pt\hbox{$\ 0\hskip.9cm 1\hskip1.0cm t-3$}}
$\bigotimes$------$\bigcirc$-- -- --
--$\bigcirc$------$\bigcirc\hskip.3cm t-2$ }}
\bigskip
\noindent
Each open node represents a string or the antipseudoparticle, while the node
labelled $\otimes$ represents the real particle.

After minimizing the free energy with respect to $P_a^+$ subject to the
constraints \lastconstr\ and \diagr\ and defining
\eqn\eddef{{\rho_a(\t)\over P_a(\t)}={\l_a e^{-\e_a(\t)}\over 1+\l_a
e^{-\e_a(\t)}},}
one finds a system of the form \TBA\ where each node in the above diagram is a
particle species (the open nodes have $m_a=0$ while the node labeled $\otimes$
is massive), and $\phi _{ab}(\t )$ is same as above.

Taking the UV limit of \fe\ involves a few tricks \ZamoI. We define
$x_a=\exp(-\epsilon_a(0))$, and $y_a=\exp(-\epsilon_a(\infty))$.
One finds that \fend\
\eqn\uvlim{c_{UV}=-{6R\over \pi}E(mR\rightarrow 0) =
{6\over \pi^2} \sum _a \left(
{\cal L}_{\lam_a}(x_a)-{\cal L}_{\lam_a}(y_a)\right),}
where
\eqn\fcftepa{{\cal L}_{\lam}(x)=\half\int_{C} d\ep\left[{\ep \lam
e^{-\ep} \over 1+\lam e^{-\ep}}+\ln(1+\lam e^{-\ep})\right],}
with the contour $C$ going from $x$ to infinity. A change of variables shows
that ${\cal L}_{1}(x)=L(x/(1+x))$, and ${\cal L}_{-1}(x)=-L(x)$, where $L(x)$
is the Rogers dilogarithm function
$$L(x)=-{1\over 2}\int_0^x dy\left[{\ln y\over (1-y)} +
{\ln(1-y)\over
y}\right],$$
It follows from
\TBA\ that the constants $x_a$ are the solutions to the equations
\eqn\xa{x_a=\prod _b (1+ \l_b x_b)^{N_{ab}},}
where $N_{ab}={1\over 2\pi}\int d\t \phi _{ab}(\t )$.  The constants $y_a$ in
\uvlim\ are nonzero only for those species $a ^{\p}$ with $m_{a^{\p}}=0$,
where they are the solutions to
\eqn\ya{y_{a^{\p}}=\prod _{b^{\p}} (1+
\l_b y_{b^{\p}})^{N_{a^{\p}b^{\p}}},}
where $b^{\p}$ also runs only over massless species.
These equations for the UV limit are true only when $\phi_{ab}=\phi_{ba}$.

We have set up the TBA formalism. We only need to specify the boundary
conditions in the $R$ direction by fixing the fugacities. If we set all
$\l_a=1$, then this corresponds to taking periodic boundary conditions: the
modified $O(n)$ model where non-contractible loops have a weight of 2.  It is
easy to find from \uvlim--\ya\ (and some dilogarithm identities \ref\Lewin{ L.
Lewin, {\it Polylogarithms and Associated Functions} (North-Holland, 1981).})
one obtains $c_{UV}=1$ as expected.

\nref\dVK{H. de Vega, M. Karowski, Nucl. Phys. B285 (1987) 619}
To get the actual $O(n)$ model TBA there are two possible approaches. The
first one is to notice that in the lattice model, non-contractible loops on
the cylinder can be given a weight $n$ if one introduces a seam \rBCN.  Using
the analogy between particle trajectories and the loops, this amounts to
computing $\hbox{Tr}\ e^{-RH}e^{2i\pi \gamma F}$ with $\gamma=1/t$, where
$F=\half(\#\hbox{solitons}-\# \hbox{antisolitons})$. More formally,
this relation follows from the fact that $\tr\ e=n$ for the original $O(n)$
representation while $\tr\ e=2$ in the sine-Gordon replacement, even though
both obey the same Temperley-Lieb algebra. {}From the Bethe Ansatz
calculation outlined in the appendix, one finds \refs{\TS,\dVK}
\eqn\F{F={t\over 2}\int d\tht (\rho_{1^-}(\tht)-\rho_{(t-1)}(\tht)),}
The appropriate trace is therefore reproduced by choosing the fugacities
\eqn\fug{\l_{1^-}=e^{-i\pi \gamma t},\l_{(t-1)}=e^{i\pi \gamma t},}
while $\l=1$ for the other species. For general $\gamma$, this amounts to
giving non-contractible loops a weight of $2\cos\gamma\pi$.

In the UV limit, one finds, using \uvlim--\ya\ (we end up with dilogarithms at
complex argument) and taking the second derivative of $c_{UV}$ with
respect to $\gamma$
\eqn\generalc{c_{UV}(\gamma)=1-6{(\gamma t)^2\over t(t+1)}.}
For the $O(n)$ model ground state, $\gamma=1/t$ so we recover \cdilute.
Notice that this choice corresponds to $\l_{(t-1)}=\l_{1^-}=-1$, while $\l=1$
for the rest. This ends up removing the nodes $t-2$, $t-1$ and $1^-$. Indeed,
for these nodes, the solutions of \xa\ are the same as \ya:
$x_{(t-1)}=y_{(t-1)}=x_{1^-}=y_{1^-}=1$ for the two end ones, and
$x_{(t-2)}=y_{(t-2)}=0$ for the node joining them. When $x_a=y_a$,
$\e_a(0)=\e_a(\infty)$, which in fact means that $\e_a(\tht)$ does not depend
on $\tht$. Thus $\e_{(t-2)}(\tht)=\infty$ for all $\tht$, it no longer
contributes to the equation for $\e_{(t-3)}$, and we see that these three nodes
decouple from the rest of the system without contributing to anything. Thus
the $O(n)$ TBA system at $n=2\cos(\pi/t)$ is described by
\bigskip
\noindent
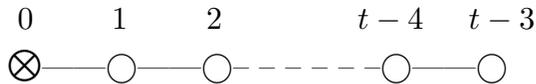
\centerline{\hbox{\rlap{\raise15pt\hbox{$\ \  0\hskip.87cm 1\hskip.87cm
2\hskip1.5cm t-4\hskip.5cm t-3$}}
{\raise1pt\hbox{$\bigotimes$}}------$\bigcirc$------$\bigcirc$-- -- --
--  -- --$\bigcirc$------$\bigcirc$}}
\bigskip
\noindent where all the fugacities $\l_a=1$.

\nref\Zamunpub{A.B. Zamolodchikov, October 1989 preprint}
Although the $O(n)$ model at $t$ integer is not identical to the $t$-th
perturbed unitary minimal model \Dfsz , their ground states coincide.  Thus we
have proven that the minimal-model TBA is described by the above diagram, as
conjectured (and proven for $t=3,4$) in \Conj.  Another way of deriving the
TBA for the unitary minimal models would be to consider a scattering theory
that uses the RSOS S-matrix \refs{\Zamunpub,\Leclair} instead of the
sine-Gordon S-matrix \Conj. Again, as was discussed in the context of lattice
models \PasSal\ or integrable perturbations \ref\Reshe{N.Y.Reshetikhin,
F.A.Smirnov, Comm. Math.  Phys. 131 (1990) 157}, the S-matrix obeys another
representation of the Temperley-Lieb algebra, so the thermodynamics deduced
from these two cases are equivalent
\foot{This is true in the regime we are interested in, but not always:
see \ref\Sal{H.Saleur, Nucl. Phys.  B360 (1991) 219}} provided one
considers the
sine-Gordon or XXZ system with twisted boundary conditions. The mere effect of
this is to forbid the $1^-$ antipseudoparticle and the $(t-1)$ string
\ref\BR{V.V.  Bazhanov and N.Y.  Reshetikhin, Int. J.  Mod. Phys. A4 (1989)
115}. The $(t-2)$ string decouples too and we are back to the above
discussion.

The mass $m$ in TBA depends on the $O(n)$ lattice parameters as
$m=({K_c-K\over sK_c})^{(t+1)/ 4}$, where $s$ is a non-universal scale.
Notice that for $\gamma=0$ the same TBA equations would describe the ground
state energy of say the 8-vertex model. In that case however, if $\delta$ is
the Boltzmann weight of the additional vertices that do not conserve $U(1)$
charge, the mass $m$ would scale as $\delta^{t/2}$, i.e.\ with a different
exponent.

We now discuss in more detail the case $n=0$, which corresponds to $t=2$. As
is well known this describes the physics of polymers: for small $n$ we the
partition function \Onpartition\ allows only a single loop, the polymer
\ref\Gennes{P.G.de Gennes, {\it Scaling concepts in polymer physics}, Cornell
University Press, (Ithaca, 1985)}. In that case the algebraic properties of the
S-matrix \foot{The S-matrices we refer to here differ from the one in \ZamoII\
by action of the permutation operator} in \ZamoII\ can also be obtained by
acting in a $Z_2$ graded space, the loop cancellation occurring here due to
boson-fermion symmetry. The minimal such S-matrix is equivalent, up to another
gauge transformation, to the above sine-Gordon S-matrix.

As discussed in \SalI\ the continuum limit of polymers or $O(n=0)$ model is
described by (twisted) $N$=2 supersymmetry \ref\EY{T.Eguchi, S.K.Yang, Mod.
Phys. Lett. A5 (1990) 1693}. This is true only when $n$ is exactly $0$, so only
properties which can be described precisely at this point are related to
supersymmetry. The thermal perturbation $\Phi_{13}$ corresponds in the $N=2$
language to the top component of the chiral primary field
$X$ \ref\LVW{W.Lerche, C.Vafa, N.P.Warner, Nucl. Phys. B324 (1989) 427} ,
hence it preserves the $N$=2 supersymmetry. Indeed, the S-matrix for this
model is precisely the sine-Gordon S-matrix for $t=2$ discussed above
\nref\rBL{D. Bernard, A. LeClair, Phys. Lett. 247B (1990) 309}
\refs{\rBL,\FI}.

The diagram describing TBA in this case has only 3 nodes, one of which is
massive. With $\gamma=0$ the ground state energy is minus the logarithm of the
largest eigenvalue of a polymer transfer matrix that describes
non-contractible loops only, with fugacity 2
\foot{Because the lattice bulk free energy vanishes in the
polymer case, we can directly identify $E$ in the numerical study of
$c_{\hbox{eff}}$ and $d$}.  The values of $c_{\hbox{eff}}(mR)$ can be
extracted from numerical solution of the TBA equations. They can also be
measured on the lattice by numerical diagonalization of the polymer transfer
matrix. We considered the honeycomb lattice with cylinders of radius up to
$R=6$ honeycomb faces.  Agreement between TBA and lattice computations was
found to be excellent, even though $s$ must be determined numerically before
most comparisons can be made.  The corresponding curve is given in
\fig\curves{Lattice measurements for the various scaled ground states
$c_{\hbox{eff}}$: non-contractible polymers only, each with a weight 2 . $d$:
one non contractible polymer.  $dd$ (c): one polymer, contractible or not.}
for $R=6$.  The main feature of this curve is the inflexion point at which the
value of $c_{\hbox{eff}}$ extracted from TBA is .7714.  Lattice estimates of
this value are
\settabs 6 \columns
\+R=&2&3&4&5&6\cr
\+&.732&.748&.754&.758&.762\cr
\noindent
which converge nicely to their expected limit. Although $c_{\hbox{eff}}$ is
not very interesting as a geometrical object, we think this numerical check is
useful. In particular the fact that thermal perturbation in polymers does not
break $N$=2 supersymmetry, although reasonable, is a little difficult to derive
from lattice analysis (due in particular to twisting) and is indirectly
checked here.

In the perturbed N=2 theory, a very interesting quantity is $\Tr\
e^{-RH}F(-)^F$. It can of course be computed from the above by introducing a
fugacity $\gamma$ close to $1/2$ and taking a derivative with respect to
$\gamma$, hence appearing as a solution of an integral equation. What is
remarkable however is that, as discussed in
\ref\Vafa{S.Cecotti, P.Fendley, K.Intriligator, C.Vafa,
``A New Supersymmetric Index'' preprint SISSA 68/92/EP, BUHEP-92-14,
HUTP-92/A021.}, it is a kind of supersymmetric index, and in this case is
related to the solution of a simple differential equation of Painleve III
type, the radial part of the sinh-Gordon equation. Introduce the function $u$,
the solution of
\eqn\painleve{u_{zz}+z^{-1}u_z=\sinh u}
where $u$ has no singularity on the real axis and has asymptotic behavior
as $z\rightarrow 0$
\eqn\boundary{u(z)\sim -2/3 \ln(z)+ \hbox{constant}}
Introduce on the other hand the generating function
\eqn\onepolymer{{\cal Z}_{nc}=\sum K^{\# \hbox{monomers}}}
where the sum is taken over all configurations of a {\bf single
non-contractible loop} on a cylinder of length L and circumference R.  One
sees from \Onpartition\ and the identification of fermion numbers for the
$O(n)$ model and the sine-Gordon theory already discussed that this is
precisely $\Tr e^{-RH}F(-)^F$. The results of \Vafa\ then give
\eqn\polynumber{{\cal Z}_{nc}(z=mR)\rightarrow -{L\over 2R}z{d\over
dz}u(z)\quad \hbox{as}\ L/R\rightarrow \infty.}
In particular, at the critical point one has ${\cal Z}_{nc}\sim {L\over 3R}$.
The differential equation \painleve\ is easy to solve numerically , and for
the lattice quantity it can be evaluated using a simple modification of the
transfer matrix used in the previous problem. Introduce
\eqn\d{d=
\hbox{Lim}_{L/R\rightarrow \infty}\ {6R\over \pi L}{\cal Z}_{nc};}
$d$ goes to $2/\pi$ in the UV limit and goes to zero in the IR.
Lattice measurements for $R=5$ are represented in \curves. Once again,
agreement with \polynumber\ is very good. For instance, the value of
$d$ at the inflexion point is .5104. Lattice estimates are
\settabs 5 \columns
\+R=&2&3&4&5\cr
\+&.492&.498&.502&.504\cr
The knowledge of \onepolymer\ gives access to very non-trivial geometrical
information. Call $\mu=1/K_c$ the effective connectivity constant for a given
lattice. Then the number of configurations of a non-contractible loop made of
$N$ monomers on a cylinder of circumference $R$ and length $L$ is expected to
scale as
\eqn\omeg{\omega_N(R)\sim LR^{-7/3}\mu^N sf(Ns/R^{4/3})}
where $f$ is a universal function and $s$ is the lattice-dependent scale
factor.  {}From the above one finds
\eqn\laplace{d(mR)={6\over \pi}\int e^{(mR)^{4/3}y}f(y)dy}
where $m=({K_c-K\over sK_c})^{3/4}$. The scaling function $f$ is just simply
obtained by the Laplace transform of the solution of the Painleve equation.

Another interesting geometrical quantity
is the generating function ${\cal Z}$ defined in a way similar to
\onepolymer\ but summing also over configurations where the loop is
contractible. To obtain the asymptotic behavior of this function is
considerably more involved. Indeed, while non-contractible loops have a
behavior fully determined by the $N$=2 theory, contractible ones depend on
properties of the $O(n)$ model for $n$ close but not strictly equal to 0. We
therefore have to discuss how to generalize the above analysis to the case
where $t$ is close to $2$. We shall restrict to
\eqn\fort{t=2+{1\over M}.}
The analysis is easier when $t$ is rational, so we take $M$ to be an integer.
This corresponds to the non-unitary minimal models ($2M+1,3M+1$).  Relying on
the analysis of \TS\ we find that the free energy is as in \fe, and the
integral equations are of the form \TBA. There is one massive particle, which
we label by $0$ as before. There are $M+2$ massless pseudoparticles. The ones
labelled $a=1,...,M,M+1$ are $|2a-1|$-strings (antistrings) if $a$ is odd
(even), while $\overline{M+1}$ is a 2-string. It turns out to be simpler to
write out the TBA integral equations \TBA\ in a form where $\phi_{12}$ is not
symmetric.  Then we have
\eqn\phiab{\eqalign{\phi_{01}(\tht)&={1\over \cosh\tht}\cr
\phi_{11}(\tht)&=\int dk e^{ik\tht}\half{\cosh((1-{1\over M}){\pi k\over
2})\over \cosh {\pi k\over 2} \cosh {\pi k \over 2 M}}\cr
\phi_{12}(\tht)=-\phi_{21}(\tht)&={M\over \cosh M\tht}\cr
\phi_{j,j+1}(\tht)=\phi_{M,\overline{M+1}}(\tht)&={M\over \cosh M\tht}\cr}}
for $j=2,\dots,M$.  All the $\phi_{ab}$ other than $\phi_{12}$ are symmetric,
and all the remaining ones are zero. This system has an incidence diagram
similar to that at $t$ integer, although here particle $1$ couples to itself.

The system with all $\l=1$ gives of course $c_{UV}=1$ for all $M$, since it
corresponds to sine-Gordon at different couplings.

To get the actual ground state energy (which is not the central charge) of the
non-unitary minimal models, we change the fugacity for the two end nodes, just
like in the unitary minimal models, by setting
$\l_{M+1}=\l_{\overline{M+1}}=-1$, while the rest remain $1$. This again has
the effect of merely removing the three end nodes ($M,M+1,\overline{M+1}$). We
have checked that the correct value $c_{UV}=1-6/(2M+1)(3M+1)$ is obtained in
that case.

To observe the actual central charge of the minimal model one needs to project
out the negative-dimension ground state. To do this, we set also $\l_1=-1$. As
before the three end nodes decouple. Let us give some details of the UV
calculation. Our choice of $\phi_{ab}$ is not symmetric; symmetric equations
are found by rewriting the equations in terms of $-\e_2,-\e_3,...  -\e_{M-1}$.
(We do not change the sign on $\e_0$ or $\e_1$.)  The UV free energy is then
proportional to
\eqn\newuv{L\left({x_0\over 1+x_0}\right)-[L(x_1)-L(y_1)]+
\sum_{j=2}^{M-1} \left[L\left({1\over
1+x_j}\right) - L\left({1\over 1+y_j}\right)\right].}
The solutions of \xa\ and \ya\ are
\eqn\xaya{\eqalign{x_0&={\sin{\pi\over 3M+1}\over \sin{M\pi\over 3M+1}}\cr
1-x_1=\left({\sin{\pi\over 3M+1}\over \sin{M\pi\over
3M+1}}\right)^2&\qquad\qquad
1-y_1=\left({\sin{\pi\over 2M+1}\over \sin{M\pi\over 2M+1}}\right)^2\cr
1+x_j=\left({\sin{(M+1-j)\pi\over 3M+1}\over \sin{\pi\over
3M+1}}\right)^2&\qquad\qquad
1+y_j=\left({\sin{(M+1-j)\pi\over 2M+1}\over \sin{\pi\over 2M+1}}
\right)^2\cr}}
We checked numerically that these give the correct answer
$c_{UV}=1-6M^2/(2M+1)(3M+1)$.  In the $M\rightarrow\infty$ limit, the
contribution from $x_0$ is order $1/M$, while the contributions of the others
are individually $1/M^2$. However, there are order $M$ such contributions, so
these also must be included.  By replacing the sum over $j/M$ with an integral
one checks directly that $c_{UV}\sim 5/6M$ as $M\rightarrow\infty$ and
therefore ${\cal Z}\sim {5L\over 18R}$.

We can recover the $t=2$ case ($M=\infty$) by noticing that except for
$\phi_{0 1}$, all the non-zero $\phi_{ab}=\pi\delta(\tht)$ at $M=\infty$. Then
all the equations \TBA\ except for $a=0,1$ are not integral equations any
more, but difference equations in $M$. As detailed in \TS, solving these gives
the $t=2$ result from above. The leading correction involves an infinite
number of coupled integral equations, which we have not been able to simplify
noticeably. Numerical solution for finite $M$ and extrapolation to $M$ infinite
produces results in good agreement with the lattice computation.
Results in the latter case for
\eqn\dd{dd=\hbox{Lim}_{L/R\rightarrow\infty}{6L\over \pi R}{\cal Z}}
are represented in \curves\ for a cylinder of R=5 honeycomb faces\foot{In this
case a non-universal bulk free energy term must be subtracted to get $E$.}.

The flow from the critical to the dense $O(n)$ model should also present
interesting features, but the scattering theory is not so well-understood and
we postpone most of its study to a subsequent publication. We shall content
ourselves with the following observation.  For unitary theories, one can
define a $c$-function, which is the central charge at the conformal point and
which never increases under renormalization group flows (increasing $mR$)
\ref\cth{A.B.Zamolodchikov, JETP Lett. 43(1986) 730}. It has been
conjectured several times that the ground-state energy on a cylinder plays the
role of an alternate $c$-function, even for non-unitary theories. The dilute-
and dense-polymer theories are non-unitary. Moreover, in the sector where
non-contractible loops get a weight 2, they both have $c_{\hbox{eff}}=1$ in
the UV. If a $c_{\hbox{eff}}$-theorem held, $c_{\hbox{eff}}(mR)$ here should
be a constant in the flow from dilute to dense.  We have estimated this
quantity numerically using our polymer simulation. As always, in the dense
region the convergence is a little worse than in the dilute region. But the
main features of the resulting curve proved very stable, with uncertainties of
only around 1\% for the asymptotic quantities.  The curve for $R=6$ hexagonal
faces is given in \fig\dense{Flow of $c_{\hbox{eff}}$ to the dense phase.}
. It violates the $c_{\hbox{eff}}=1$ theorem in a
spectacular fashion. Moreover, the maximum and minimum lie respectively very
close to 81/70 and 7/10. These are the two values of the central charge
neighboring $c=1$ in the minimal $N=1$ series.  Recall that the $N$=2 theory
related to polymers belongs also to the $N$=1 series.  The flow between dilute
and dense polymers seems therefore to be an example of a ``roaming
trajectory'' as introduced in
\ref\ZamoIIII{Al.B.Zamolodchikov, ``Resonance Scattering and Roaming
Trajectories'' Ecole Normale preprint ENS-LPS-335}.

\bigskip\bigskip\centerline{{\bf Acknowledgments}}\nobreak
We would like to thank K. Intriligator, G. Moore, C. Vafa and A.
Zamolodchikov for useful conversations. H.S. was supported by the Packard
Foundation and DOE grant DE-AC-76ERO3075, while P.F.\ was supported by DOE
grant DEAC02-89ER-40509.

\appendix{A}{Derivation of sine-Gordon TBA equations}

In this appendix, we derive the equations \lastconstr\ and \diagr\ for the
density of states of the sine-Gordon model at integer $t$.  The allowed
$\Lambda(\tht_i|\tht _1,\tht_2 \dots \t _N)$ in \dos\ are the eigenvalues of
the sine-Gordon ``transfer matrix'' $T_{ab}(\tht_i|\tht _1,\tht_2 \dots \t
_N)$ for bringing the $i$-th particle (with rapidity $\tht_i$) through the
other $N-1$ particles and ending up with itself. The components of
$T_{ab}(\tht_i )$ can be written in terms of the S-matrix elements as
\eqn\Tab{(T_{ab}(\tht_i))_{\{c_j\}}^{\{d_j\}}\equiv
\sum_{\{k_j\}=0,1} S_{k_Nc_{1}}^{d_{1}k_{1}}(\tht -\tht_{1})
S_{k_{1}c_{2}}^{d_{2}k_{2}}(\tht-\tht_{2})\dots
S_{k_{N-1}c_{N}}^{d_{N}k_N}(\tht-\tht_{N}),}
where the $c_j$ $(d_j)$ label whether each of the $N$
particles before (after) the scattering process is a soliton or antisoliton;
the $k_j$ are the intermediate particles.
The sine-Gordon S-matrix is given in \Sinegordon, where
\eqn\forZ{Z(\tht)=\exp\left({i\over 2}
\int_{-\infty}^{\infty} {dk\over k} {\sin (\tht k)} {\sinh (t-1){\pi k\over 2}
 \over \sinh t{\pi k \over 2}\cosh {\pi k\over 2}}\right)}

Formally the sine-Gordon ``transfer matrix'' is equivalent to the transfer
matrix of the six-vertex model or of the XXZ spin chain. To diagonalize
it, we can rely on the well-discussed technique of the Bethe ansatz \ref\rBA{
see, for example, L. Faddeev or J. Lowenstein, in Les Houches 1982 {\it Recent
advances in field theory and statistical mechanics} (North-Holland 1984),
edited by J.-B.  Zuber and R. Stora.}. The only additional ingredient here is
the prefactor $Z(\tht)$; by the definition \Tab\ we see that its presence
causes the eigenvalues $\Lambda(\tht_i)$ to be multiplied by $\prod_{j=1}^N
Z(\tht_i-\tht_j)$. The result for integer $t$ (for example, see \TS\ or \dVK,
or \FI\ for the case $t=2$), is that
\eqn\evconstr{{d\over d\tht_i}\ln\Lambda(\tht_i)=
({d\over d\tht_i}Im \ln Z) *P_0^+
- \sum_a p_a*P_a^+ + p_{1^-}*P_{1^-}^+,}
and
\eqn\XXZBA{\eqalign{2\pi P_a&=p_a*P_0^+ +\sum_b
\Theta_{ab}*P_b^+ - q_a*P_{1^-}^+,\cr
P_{1^-}&=P_{(t-1)}\cr}}
where the sums run over the strings (but not the antipseudoparticle).  The
kernels in these equations are defined most easily in terms of their Fourier
transforms:
\eqn\forphi{\tilde\Theta_{ab}(k)=
\int {d\tht\over 2\pi} e^{ik\tht}\Theta_{ab}(\tht)=
\delta_{ab}-2{\cosh{\pi k\over 2} \sinh(t-a){\pi k\over 2}
\sinh b {\pi k\over 2} \over \sinh {\pi k\over 2}\sinh t{\pi k\over 2}}}
for $a\geq b$ with $\Theta _{ab}=\Theta _{ba}$, and
\eqn\forqp{\eqalign{\tilde p_a(k)&={\sinh (t-a){\pi k\over 2}\over\sinh t{\pi
k\over 2}}\cr
\tilde p_{1^-}&=\tilde p_{(t-1)}\cr
\tilde q_{b}&=\tilde\Theta_{(t-1),b} \cr}}

To simplify these equations, it is much more convenient to work in Fourier
space. We define $\tilde z(k)$ as the Fourier transform of ${d\over d\tht_i}Im
\ln Z $, so that
$$2\tilde z(k) \cosh {\pi k\over 2}= {\sinh (t-1){\pi k\over 2} \over \sinh
{\pi k \over 2}t}=\tilde p_1$$
It is easy to prove the following trigonometric
identities:
\eqn\trigid{\eqalign{2 \tilde\Theta_{ab}\cosh{\pi k\over 2}&=
\tilde\Theta_{a+1,b} + \tilde\Theta_{a-1,b} - \delta_{a+1,b} -
\delta_{a-1,b}\cr
2\tilde p_a \cosh{\pi k\over 2}&=\tilde p_{a-1}+\tilde p_{a+1}\cr
2\tilde q_a\cosh{\pi k\over 2}&=\tilde q_{a-1}+\tilde q_{a+1}
-\delta_{a,t-2}-\delta_{a,t}
.\cr}}
Using these in \XXZBA\ and \evconstr\ gives
\eqn\finalP{\eqalign{2\tilde P_a \cosh{\pi k\over 2} - \tilde P_{a+1}-
\tilde P_{a-1}&= -\tilde P_{a+1}^+ -\tilde P_{a-1}^+ +
\delta_{a,t-2}P_{1^-}^+ \cr
2\tilde P_{(t-1)} \cosh{\pi k\over 2} -\tilde P_{(t-2)}&=
-\tilde P_{(t-2)}^+\cr
2\tilde l(k) \cosh{\pi k\over 2} -\tilde P_{1}&=-\tilde P_{1}^+.\cr}}
where $\tilde l(k)$ is the Fourier transform of ${d\over d\tht}Im
\ln \Lambda(\tht)$, and $a=1,\dots,t-2$.
Using $P^-\equiv P-P^+$ and the fact that the Fourier transform of
$1/(2\cosh{\pi k\over 2})$ is $1/\cosh\tht$ gives us \lastconstr\ and \diagr.

{}From the Bethe ansatz one also finds that the number of solitons minus the
number of antisolitons is
\eqn\forF{\int d\tht \left(-P_0^+(\tht)
+2\sum_a a P_a^+(\tht) +2P_{1^-}^+(\tht)\right)}
In the XXZ language, this is twice the value of the third component of the
spin.  Using the relation \XXZBA\ for $a=p-1$ gives the relation \F.

 \listrefs \listfigs \bye